\documentclass[aps,prl,twocolumn,showpacs,preprintnumbers,floatfix]{revtex4-1}

\usepackage{color}
\usepackage{graphicx}% Include figure files
\usepackage{bm}% bold math

\begin{document}

\title{Ultracold hydrogen atoms: a versatile coolant to produce ultracold
molecules}

\author{Maykel L. Gonz\'alez-Mart\'inez}
\altaffiliation{Present address:
             Laboratoire Aim\'e Cotton, CNRS,
             Universit\'e Paris-Sud XI,
             ENS Cachan, B\^{a}t.\ 505, Campus d'Orsay,
             91405 Orsay, France}
\email{maykel.gonzalez-martinez@u-psud.fr}
\affiliation{Joint Quantum Centre (JQC) Durham/Newcastle,
             Department of Chemistry,
             Durham University, South Road,
             Durham DH1~3LE, United Kingdom}
\author{Jeremy M. Hutson}
\email{J.M.Hutson@durham.ac.uk}
\affiliation{Joint Quantum Centre (JQC) Durham/Newcastle,
             Department of Chemistry,
             Durham University, South Road,
             Durham DH1~3LE, United Kingdom}
% \affiliation{Joint Quantum Centre (JQC)
% Durham/Newcastle, Department of Chemistry, Durham University, South Road,
% Durham, DH1~3LE, United Kingdom}

\date{\today}

\begin{abstract}
We show theoretically that ultracold hydrogen atoms have very favorable
properties for sympathetic cooling of molecules to microkelvin temperatures. We
calculate the potential energy surfaces for spin-polarized interactions of H
atoms with the prototype molecules NH($^3\Sigma^-$) and OH($^2\Pi$) and show
that they are shallow (50 to 80 cm$^{-1}$) and only weakly anisotropic. We
carry out quantum collision calculations on H+NH and H+OH and show that the
ratio of elastic to inelastic cross sections is high enough to allow
sympathetic cooling from temperatures well over 1~K for NH and around 250 mK
for OH.
\end{abstract}

\maketitle

Samples of ultracold molecules, at temperatures below 1 millikelvin, have
potential applications in fields ranging from high-precision measurement to
quantum simulation \cite{Carr:NJPintro:2009}. Ultracold polar molecules are
particularly interesting because they have long-range anisotropic interactions
whose strength can be tuned with applied electric field, which are important in
understanding the properties of dipolar quantum gases.

There have been great advances in producing ultracold molecules in the last few
years. Many groups have succeeded in producing alkali-metal dimers in
high-lying vibrational states by either magnetoassociation or photoassociation
\cite{Hutson:IRPC:2006, Kohler:RMP:2006, Jones:RMP:2006}, and a small number of
such species have been transferred to their absolute ground states, either
incoherently by absorbtion followed by spontaneous emission \cite{Sage:2005,
Deiglmayr:2008} or coherently by stimulated Raman adiabatic passage (STIRAP)
\cite{Lang:ground:2008, Ni:KRb:2008, Danzl:ground:2010, Aikawa:2010}. KRb
molecules produced by STIRAP \cite{Ni:KRb:2008} have been used to investigate
ultracold chemical reactions \cite{Ospelkaus:react:2010, Ni:2010} and the
properties of dipolar quantum gases \cite{Ni:2010}.

The indirect methods that produce ultracold molecules via ultracold atoms are
inherently limited to molecules formed from atoms that can themselves be
cooled. So far this has restricted molecule formation experiments to the
alkali-metal dimers, although there are prospects for extending the approach to
a wider range of species, such $^2\Sigma$ molecules formed from an alkali-metal
atom and an alkaline-earth \cite{Zuchowski:RbSr:2010} or Yb
\cite{Brue:LiYb:2012, Brue:AlkYb:2013} atom. However, there is also great
interest in producing ultracold samples of a wider range of species, including
polyatomic molecules such as ND$_3$ \cite{Bethlem:trap:2000} and CH$_3$F
\cite{Zeppenfeld:2012} and reactive species such as NH
\cite{vandeMeerakker:2001, Egorov:2004, Campbell:2007, Hoekstra:2007,
Riedel:2011} and OH \cite{vandeMeerakker:OH:2005, Wohlfart:2008, Hudson:2004,
Sawyer:2007, Sawyer:2008, Sawyer:2011, Stuhl:evap:2012}. Methods such as
buffer-gas cooling \cite{Weinstein:CaH:1998}, molecular beam deceleration
\cite{Bethlem:trap:2000} and velocity filtering \cite{Rieger:2005} have been
developed, which are capable of producing trapped samples at temperatures
between 10 and 500 mK, but the ultracold regime (below 1 mK) has not yet been
reached.

There is thus a great need for a second-stage cooling method that can cool
molecules from tens or hundreds of millikelvin to the ultracold regime. The
principal candidates are laser cooling, evaporative cooling, and sympathetic
cooling. Laser cooling was for a long time dismissed as a method of cooling
molecules, because electronically excited molecules can in principle reradiate
to many different vibrational levels of the electronic ground state. However,
in the last few years considerable progress towards laser cooling has been made
for molecules with unusually good overlap between the ground and excited
vibronic states \cite{Shuman:2010, Barry:beam:2012}, and Hummon {\em et al.}\
\cite{Hummon:2013} have succeeded in producing a magnetooptical trap for YO
molecules. Nevertheless, laser cooling is likely to remain applicable to only a
very restricted set of molecules.

\begin{figure*}[t]
\includegraphics[width=175mm]{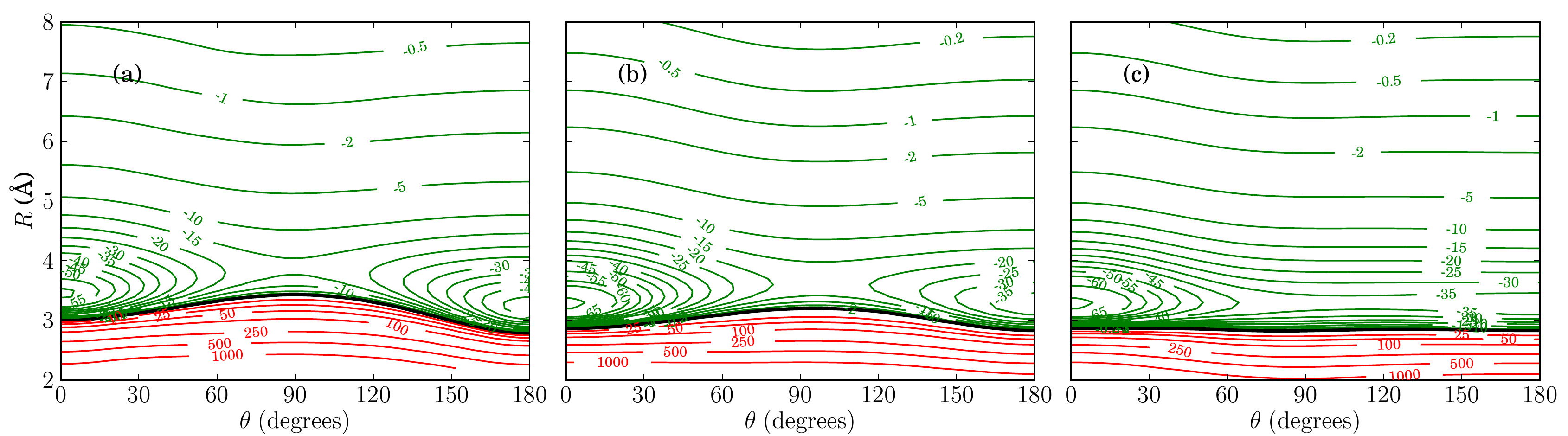}
\caption{\emph{Ab initio} interaction potentials for: (a) the $^4A''$ high-spin
state of H+NH; (b) and (c) the $^3A'$ and $^3A''$ high-spin states of H+OH,
respectively. Contours are labeled in units of cm$^{-1}$, with $180^{\circ}$
corresponding to the H atom approaching the molecule from the hydrogen side.}
\label{fig:surfaces}
\end{figure*}

Sympathetic cooling relies on the thermalization of the `warm' species of
interest by collisions with ultracold atoms. It has been widely used to cool
molecular ions (for a recent compilation, see \cite{Roth:2009}) and neutral
atoms \cite{Myatt:1997, Modugno:2001}.  Its use for neutral molecules was
proposed by Sold\'an and Hutson \cite{Soldan:2004} in 2004, but it has not yet
been experimentally demonstrated. The difficulty is that static electric and
magnetic traps can confine molecules only when they are in
\emph{low-field-seeking} states, and these states are never the lowest state in
the applied field \footnote{An alternative proposal for sympathetic cooling has
been made by Barletta {\em et al.}\ \cite{Barletta:2010}, who propose cooling
of molecules such as H$_2$ and C$_6$H$_6$ with ground-state rare-gas atoms held
in an optical trap. However, even the extremely high (intra-cavity) laser
intensities proposed in ref.\ \cite{Barletta:2010} produce a trap depth of only
25~mK for Ar and 150~mK for C$_6$H$_6$. Loading of either ground-state atoms or
molecules into such traps has not yet been achieved.}. Collisions that transfer
molecules to the lower states release kinetic energy and eject both the atoms
and the molecules from the trap. The key quantity that determines the
feasibility of sympathetic cooling is the ratio $\gamma$ between the cross
section for elastic collisions (which produce thermalization) and that for
inelastic collisions (which cause trap loss). A common rule of thumb is that,
for cooling to be successful, this ratio needs to be at least 100
\cite{deCarvalho:1999}.

There has been a long search for atom-molecule pairs that would be good for
sympathetic cooling. However, extensive theoretical work
\cite{Wallis:MgNH:2009, Gonzalez-Martinez:hyperfine:2011, Wallis:LiNH:2011,
Tscherbul:poly:2011, Parazzoli:2011} and some experimental work
\cite{Parazzoli:2011} has shown that, for most experimentally accessible
combinations of atoms and molecules, inelastic collisions will lead to
unworkable trap losses. Inelastic cross sections can be suppressed at low
collision energies by \emph{centrifugal barriers} that exist for the collision
products \cite{Avdeenkov:2001, Volpi:2002, Wallis:MgNH:2009,
Gonzalez-Martinez:hyperfine:2011, Wallis:LiNH:2011}, but even for light atoms
such as lithium the barriers are only around 3 to 12~mK high
\cite{Wallis:LiNH:2011}, and strong inelasticity sets in for collisions above
this energy. The best system proposed so far is Mg+NH \cite{Wallis:MgNH:2009,
Gonzalez-Martinez:hyperfine:2011}, where sympathetic cooling is predicted to
succeed if the molecules can be precooled to 10-20 mK. Unfortunately, such
temperatures can so far be achieved only for very small numbers of molecules.

The purpose of the present Letter is to propose {\em ultracold atomic hydrogen}
as a versatile sympathetic coolant for molecules. Magnetically trapped hydrogen
atoms have been produced at temperatures of 40 to 100 mK and densities up to
$3\times10^{14}$ cm$^{-3}$ by purely cryogenic methods \cite{Hess:1987,
vanRoijen:1988}, and then evaporatively cooled to produce a Bose-Einstein
condensate (BEC) of $10^9$ atoms at a temperature around 50 $\mu$K and
densities between $10^{14}$ and $5\times10^{15}$ cm$^{-3}$ \cite{Fried:1998}.
For sympathetic cooling purposes BEC is unnecessary, but the large densities
and cloud sizes achievable are very valuable. Furthermore, because of the low
mass and small polarizability of atomic hydrogen, the centrifugal barriers for
collisions with molecules such as NH and OH are around 400 mK high. As will be
seen below, this produces very favorable conditions for sympathetic cooling,
starting from temperatures of 250 mK for OH and over 1~K for NH. This is a
crucial improvement over earlier proposals, because very large numbers of cold
molecules can be produced at these higher temperatures.

We consider NH($^3\Sigma^-$) and OH($^2\Pi$) as prototype molecules. NH
molecules have been cooled by buffer-gas cooling \cite{Egorov:2004} and trapped
at a peak density of $10^8$~cm$^{-3}$ and a temperature of a few hundred mK
\cite{Campbell:2007}. NH has also been decelerated \cite{vandeMeerakker:2006}
and trapped electrostatically \cite{Hoekstra:2007} in its $a{^1\Delta}$ state
and subsequently pumped into the $^3\Sigma^-$ state \cite{Riedel:2011},
allowing accumulation in a magnetic trap. OH($^2\Pi$) has also been decelerated
electrostatically \cite{Hudson:2004, vandeMeerakker:OH:2005, Hudson:2006,
Wohlfart:2008} and trapped both electrostatically \cite{vandeMeerakker:OH:2005}
and magnetically \cite{Sawyer:2007, Stuhl:dyn:2012}, and Stuhl {\em et al.}
\cite{Stuhl:evap:2012} have very recently achieved evaporative cooling to a
temperature around 5~mK.

\begin{figure*}[t]
\includegraphics[width=178mm]{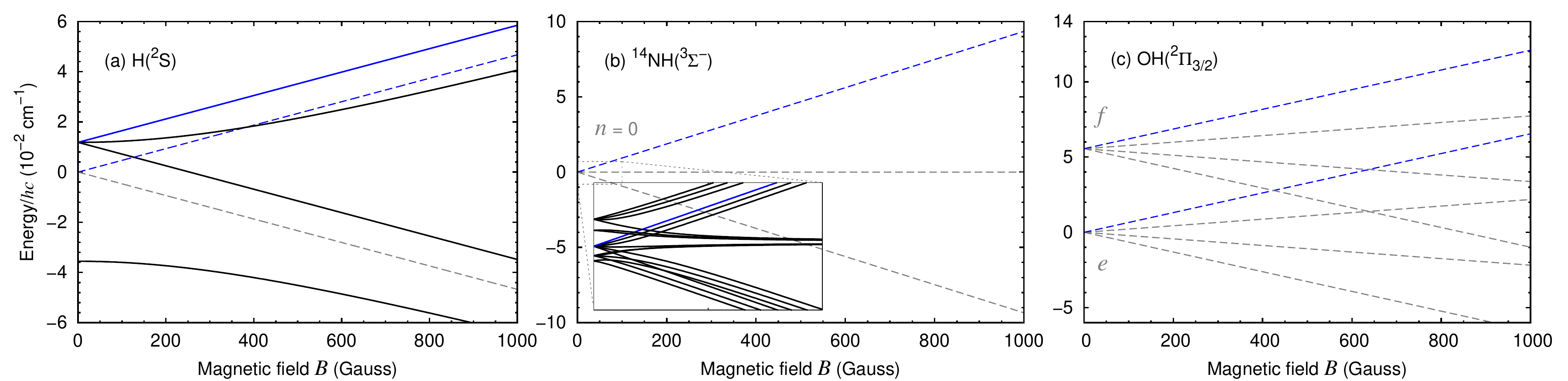}
\caption{Magnetic-field dependence of the energy levels for: (a) H($^2$S); (b)
$^{14}$NH($^3\Sigma^{-}$); and (c) OH($^2\Pi_{3/2}$). Solid (dashed) lines
correspond to the inclusion (exclusion) of hyperfine terms. The magnetically
trappable states for which scattering calculations were carried out are shown
in blue.} \label{fig:elevels}
\end{figure*}

We have calculated the potential energy surfaces for interaction of
ground-state NH($^3\Sigma^-$) and OH($^2\Pi$) molecules with H atoms. For NH
there are two surfaces, of $^2A''$ and $^4A''$ symmetry, whereas for OH the H
atom splits the degeneracy of the $\Pi$ state to produce 4 surfaces, of $^1A'$,
$^1A''$, $^3A'$ and $^3A''$ symmetry. Our primary interest is in collisions of
atoms and molecules in spin-stretched states, in which all the quantum numbers
for the projections of angular momenta onto the magnetic field direction have
their maximum values. Such collisions occur principally on the {\em high-spin}
potential surfaces. By contrast with systems like NH+NH \cite{Janssen:2013},
there are no exothermic reaction channels on the low-spin surfaces. We have
calculated the high-spin surfaces using an open-shell restricted version of the
coupled-cluster method with single, double and noniterative triple excitations,
RCCSD(T), as implemented in MOLPRO \cite{MOLPRO_brief:2006}.
Correlation-consistent aug-cc-pVQZ basis sets \cite{Dunning:1989} were used,
with spdf functions for H and spdfg functions for N and O. The basis sets also
included spdf bond functions half-way between the molecule and the H atom, as
in ref.\ \cite{Soldan:MgNH:2009}. All interaction energies were corrected for
basis-set superposition error using the counterpoise correction
\cite{Boys:1970}. The resulting surfaces are shown in Fig.\ \ref{fig:surfaces},
and it may be seen that they are fairly weakly anisotropic, with well depths
between 60 and 80 cm$^{-1}$ and anisotropies of tens of cm$^{-1}$. This may be
contrasted with the interaction potentials with alkali-metal atoms, which for
NH have anisotropies ranging from 500 to 1700 cm$^{-1}$
\cite{Soldan:MgNH:2009}. The NH and OH bond lengths were fixed at the
equilibrium for the free molecules, 1.95770~$a_0$ and 1.83417~$a_0$. This
prevents exothermic reactions to form N($^4$S)+H$_2$ or O($^3$P)+H$_2$, but in
each case there is a substantial barrier to reaction (1000~K for NH
\cite{Zhai:2011} and 5200~K for OH \cite{Rogers:2000}) that is likely to
suppress reactive processes in the ultracold regime.

To evaluate the feasibility of sympathetic cooling, we need to calculate the
ratio $\gamma$ of elastic to inelastic cross sections as a function of
collision energy $E$ and magnetic field $B$. As a first step, we carried out
coupled-channel calculations of H+NH($^3\Sigma^{-}$) and H+OH($^2\Pi_{3/2}$)
collisions, neglecting hyperfine interactions, using the MOLSCAT program
\cite{molscat:v14-short} with computational methods identical to those in
previous work on N+NH \cite{Zuchowski:NNH:2011} and N+OH
\cite{Skomorowski:NOH:2011} collisions. The methods used are described briefly
in the Supplemental Material \cite{sup-mat-H+mol}.

The energy levels of H, NH and OH are shown as a function of magnetic field $B$
in Fig.\ \ref{fig:elevels}. We have calculated elastic and total inelastic
cross sections for atoms and molecules initially in the magnetically trappable
states shown as dashed blue lines. Inelastic collisions that change the state
of either the atom or the molecule are fully included, but those that change
the molecular state dominate. Incoming partial waves up to $L=2$ are included
to give convergence of cross sections up to collision energies of about 4~K;
the $L=3$ centrifugal barriers are about 6~K high. These calculations set the
low-spin surface $V_{^2A''}$ equal to $V_{^4A''}$. This latter approximation is
reasonable because states with different total spin $S$ are coupled only by
weak monomer fine-structure and hyperfine terms.

\begin{figure*}
\includegraphics[width=175mm]{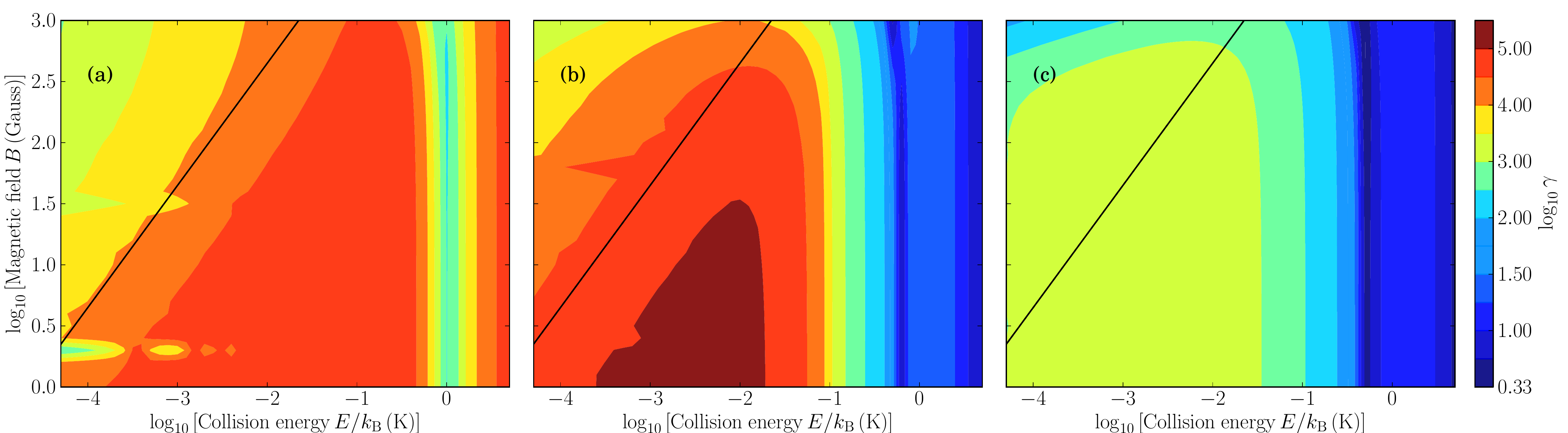}
\caption{Contour plots of the ratio $\gamma$ of elastic to inelastic cross
sections as a function of collision energy and magnetic field.  The panels
correspond to collisions of spin-polarized hydrogen with: (a) NH; and (b)
OH($e$); (c) OH($f$). The initial states are shown as dashed blue lines in
Fig.~\ref{fig:elevels}.} \label{fig:contour}
\end{figure*}

The ratio $\gamma$ is shown as a contour plot in Figure \ref{fig:contour}(a),
for H+NH($^3\Sigma^-$). The solid black line shows the maximum magnetic field
sampled in a quadrupole trap with an energy of $6k_{\rm B}T$, chosen so that
99.9\% of molecules sample fields below the line. It may be seen that $\gamma$
remains above $10^4$ for almost all energies and fields below the line at
temperatures up to well above 1~K, except for a fairly narrow band around 1~K
where $\gamma$ decreases to about 100; this is due to a d-wave resonance in the
incoming channel. The results in Fig.\ \ref{fig:contour}(a) suggest that
sympathetic cooling can succeed for NH molecules at remarkably high initial
temperatures: even the resonance near 1~K can probably be crossed in a few
collisions with relatively little loss of molecules from the trap.

We have investigated the robustness of this result to the approximations made
in Fig.~\ref{fig:contour}(a). First, we have investigated the effect of scaling
the potential energy surface by $\pm5$\% from the one shown in Fig.\
\ref{fig:surfaces}(a). The resonance around 1~K shifts by about $\mp0.2$~K over
this range but the value of $\gamma$ close to the resonance does not change
significantly. The dependence on the potential is far weaker for collisions
with H atoms than for collisions with heavier atoms and deeper interaction
potentials \cite{Zuchowski:NH3:2009, Wallis:MgNH:2009, Wallis:LiNH:2011}.
Secondly, we have investigated the effect of introducing a deeper well for the
$^2A''$ state. The true doublet surface is around 35,000 cm$^{-1}$ deep, and we
cannot converge our calculations with such a deep well, but we have explored
the effects of introducing an approximate $^2A''$ surface given by
$V_{^2A''}=\lambda_2 V_{^4A''}$ ($1<\lambda_2\le1.25$). Additional resonances
occur for some values of $\lambda_2$, but we did not observe the ratio $\gamma$
dropping below 200 in any of the cases we investigated. Finally, we have
investigated the effect of including the hyperfine interactions for both $^1$H
and $^{14}$NH, using a generalization of the methods developed in ref.\
\cite{Gonzalez-Martinez:hyperfine:2011} to include the atomic hyperfine term.
These calculations are too expensive to produce a full contour plot, but we
find that the main effects are the same as observed for Mg+NH
\cite{Gonzalez-Martinez:hyperfine:2011}: the existence of energy splittings
that persist at zero field (see solid lines in Fig.~\ref{fig:elevels}(a) and
inset of Fig.~\ref{fig:elevels}(b)) makes the kinetic energy release almost
independent of field below about 100~G for collisions that change the atomic
hyperfine state. The total inelastic cross sections are almost independent of
field and proportional to $E^{-1/2}$ in this region, while the elastic cross
sections are essentially constant. The ratio $\gamma$ is thus proportional to
$E^{1/2}$. For NH initially in its spin-stretched state with total spin
projection $m_f=+5/2$, $\gamma$ remains above 2300 at collision energies above
50 $\mu$K. Our overall conclusion from all these tests is that none of the
approximations made in Fig.~\ref{fig:contour}(a) will qualitatively affect the
success of sympathetic cooling.

For H+OH the situation is a little more complex. OH($^2\Pi$) exhibits
$\Lambda$-doubling, which splits its lowest rotational state $j=3/2$,
$\omega=3/2$ into $e$ and $f$ components, with the $f$ state lying about 0.056
cm$^{-1}$ above the $e$ state at zero electric field, as shown in Fig.\
\ref{fig:elevels}(c). The state that can be decelerated and trapped
electrostatically correlates with the $f$ component, but molecules in either
the $e$ or the $f$ state may be trapped magnetically in the $m_j=+3/2$
sublevel. H atoms must be trapped magnetically, so we have carried out
coupled-channel calculations for collisions of H ($m_s=+\textstyle\frac{1}{2}$)
with OH in the $m_j=+3/2$ sublevel of both the $e$ and $f$ states. The
resulting contour plots of the ratio $\gamma$ are shown in
Figs.~\ref{fig:contour}(b) and \ref{fig:contour}(c). It may be seen that the
results for the $e$ state are again very suitable for sympathetic cooling, with
$\gamma>10^4$ for energetically accessible fields at temperatures up to around
100 mK. $\gamma$ remains above 100 for temperatures up to about 500 mK. The
larger inelastic cross sections compared to H+NH reflect the fact that, for
$j=3/2$, the interaction potential can drive spin-changing collisions directly,
while for NH($^3\Sigma^-,n=0$) a higher-order mechanism is involved
\cite{Krems:mfield:2004}. For H+OH($f$), $\gamma$ does not become as large at
low fields, because of the possibility of relaxation to produce OH($e$) with a
kinetic energy release around 80 mK $\times k_{\rm B}$, but nevertheless
remains above 100 for all accessible fields at temperatures up to about 250 mK.
This contrasts with the situation for collisions of magnetically trapped
OH($f$) with other atoms \cite{Lara:PRA:2007, Skomorowski:NOH:2011}, even He
\cite{Tscherbul:faraday:2009}, where the kinetic energy release overcomes the
centrifugal barrier and inelastic collisions are too fast for cooling.

Ultracold H atoms have in the past been produced inside a cryogenic
environment. It is likely to be quite difficult to introduce a molecular sample
into such an environment. An alternative that appears preferable is to extract
the hydrogen cloud from the cryogenic environment before evaporative cooling.
Alternatively, developments in laser technology may in future allow Doppler
cooling of H atoms. In either case, the most obvious experiments are either to
make a separate molecular cloud and bring it into coincidence with the atomic
cloud, as has been done for Rb+ND$_3$ \cite{Parazzoli:2011}, or to decelerate
molecules almost to rest at the location of the atomic cloud. Our calculations
give elastic cross sections $\sigma_{\rm el}$ for NH and OH colliding with H
atoms in the range 200 to 400 \AA$^2$; the H+NH s-wave cross section varies by
only $\pm10$\% when the interaction potential is scaled by $\pm5$\%. For
$\sigma_{\rm el}=400$ \AA$^2$ and an H-atom sample initially at $T=50$ $\mu$K
and density $n_{\rm H}=3\times10^{14}$ cm$^{-3}$, we estimate that cooling from
1~K to 100~$\mu$K will take 80 to 90 collisions and be achieved in about 5~s,
which is within the lifetime of the atomic cloud.

The very large densities and cloud sizes available for H atoms offer another
intriguing possibility. For the densities and cross sections above, the mean
free path $(n_{\rm H}\sigma_{\rm el})^{-1}$ for atom-molecule collisions is
around 1~mm. Higher H atom densities may be achievable, and other molecules may
well have larger cross sections. It may therefore be possible to direct a
low-energy beam directly onto an atomic cloud, without a final deceleration
stage to bring it to rest, and to rely on collisions with H atoms to remove
enough kinetic energy for the molecules to be trapped.

In conclusion, we propose that {\em ultracold hydrogen atoms} are an extremely
promising medium to achieve sympathetic cooling of a wide range of molecules.
Because of the large centrifugal barriers in collisions with H atoms, inelastic
collisions are substantially suppressed at collision energies below 200 mK, or
in some cases much higher.

The authors are grateful to EPSRC for funding and to Tim Softley and Jook
Walraven for valuable discussions about the experimental possibilities.

\bibliography{../all}
\end{document}